\documentclass{article}
\usepackage{spconf,amsfonts}
\usepackage[nolist]{acronym}
\usepackage[utf8]{inputenc}
\usepackage{scalerel}
\usepackage{amsmath}
\usepackage{bbm}
\usepackage{cite}
\usepackage{multirow}
\usepackage{dblfloatfix} 
\usepackage{float}
\usepackage{comment}
\usepackage{xcolor}

\begin{acronym}[myacro]
	\acro{AoA}{angle-of-arrival}
	\acro{BS}{base station}
	\acro{ConvLSTM}{convolutional long short-term memory}
	\acro{HBF}{hybrid beamforming}
	\acro{LoS}{line-of-sight}
	\acro{FC-LSTM}{fully connected long short-term memory}
	\acro{MIMO}{multiple-input multiple-output}
	\acro{mmWave}{millimeter-wave}
	\acro{NLOS}{non-line-of-sight}
	\acro{OFDM}{orthogonal frequency division multiplexing}
	\acro{QuaDRiGa}{Quasi Deterministic Radio Channel Generator}
	\acro{SNR}{signal-to-noise ratio}
	\acro{SSF}{small scale fading}
	\acro{UE}{user equipment}
	\acro{UPA}{uniform planar array}
\end{acronym}

\DeclareMathOperator*{\argmax}{arg\,max}

\newcommand{\nbrace}[1]{{\left ( {#1} \right )}} 
\newcommand{\cbrace}[1]{{\left \{ {#1} \right \}}} 
\newcommand{\recbrace}[1]{ \left [ #1 \right ] } 
\newcommand{\norm}[1]{{\left \| {#1} \right \|}} 
\newcommand{\abs}[1]{ \left | #1 \right | } 
\newcommand{\vect}[1]{ \mathbf{#1} } 
\newcommand{\blkdiag}[1]{ \text{blkdiag} \nbrace{ #1 } } 
\newcommand{\complex}{{\mathbb C}} 

\def\HmmkH{\mathbf{H}^H\recbrace{\bar{k}}}
\def\Hmmk{\mathbf{H}\recbrace{\bar{k}}}
\def\FRF{\mathbf{F}_{\text{RF}}}
\def\FRFplus{\widehat{\mathbf{F}}_{\text{RF}}}
\def\FRFminus{\widetilde{\mathbf{F}}_{\text{RF}}}
\def\FRFH{\mathbf{F}^H_{\text{RF}}}
\def\FDK{\mathbf{F}\recbrace{\bar{k}}}
\def\FDKH{\mathbf{F}^H\recbrace{\bar{k}}}
\def\FRFstar{\mathbf{F}^{\star}_{\text{RF}}}
\def\FDKstar{\mathbf{F}^{\star}\recbrace{\bar{k}}}

\title{Deep Learning Based Hybrid Precoding \\ in Dual-Band Communication Systems}

\name{Rafail Ismayilov$\hspace{1pt}^*$, Renato L. G. Cavalcante$\hspace{1pt}^{* \dag}$ and Sławomir Stańczak$\hspace{1pt}^{* \dag}$ \vspace{-5pt}}
\address{$^*$Fraunhofer Heinrich Hertz Institute, Berlin, Germany\\	
	$^\dag$Technical University of Berlin, Berlin, Germany \vspace{-9pt} \\}

\begin{document}
\frenchspacing
\ninept

\def\dskip{2.65pt}

\setlength{\belowdisplayskip}{\dskip}

\setlength{\abovedisplayskip}{\dskip}

\def\baselinestretch{.90}

\maketitle

\begin{abstract}
	We propose a deep learning-based method that uses spatial and temporal information extracted from the sub-6GHz band to predict/track beams in the \ac{mmWave} band. 
	In more detail, we consider a dual-band communication system operating in both the sub-6GHz and \ac{mmWave} bands.
	The objective is to maximize the achievable mutual information in the \ac{mmWave} band with a hybrid analog/digital architecture where analog precoders (RF precoders) are taken from a finite codebook.
	Finding a RF precoder using conventional search methods incurs large signalling overhead, and the signalling scales with the number of RF chains and the resolution of the phase shifters.
	To overcome the issue of large signalling overhead in the \ac{mmWave} band, the proposed method exploits the spatiotemporal correlation between sub-6GHz and \ac{mmWave} bands, and it predicts/tracks the RF precoders in the \ac{mmWave} band from sub-6GHz channel measurements.
	The proposed method provides a smaller candidate set so that performing a search over that set significantly reduces the signalling overhead compared with conventional search heuristics.
	Simulations show that the proposed method can provide reasonable achievable rates while significantly reducing the signalling overhead.
\end{abstract}
\vspace{-1pt}
\vspace{-8pt}
\section{Introduction}
\label{sec:intro}
\vspace{-7pt}
Integrated networks that enable simultaneous usage of sub-6GHz and \ac{mmWave} bands have been considered as built-in technology in 5G \cite{Semiari_2019}.
Unlike the conventional sub-6GHz band, the signal processing in \ac{mmWave} band is typically accomplished in both the analog and digital domains \cite{Heath_2016}.
One of the main challenges in these hybrid systems is that conventional channel estimation methods typically used in the sub-6GHz band cannot be straightforwardly used in mmWave systems. 
The reason is that in \ac{mmWave} systems with hybrid architectures the channel measured in the digital baseband is intertwined with the choice of analog precoders (RF precoders), and thus the entries of the channel matrix cannot be directly accessed.
As a result, new methods are required to find both the analog and digital parameters in \ac{mmWave} systems with hybrid architectures.
In particular, the channel experienced by the receiver is affected by the choice of RF precoder, which is usually selected from a finite codebook, and it includes the beams in a particular angular direction.
To find the best \ac{mmWave} beam (e.g., the angular direction of the strongest signal) the transmitter and receiver need to perform an exhaustive search over all possible beam patterns \cite{3gpp_2017}.
An exhaustive search, in turn, creates a signalling overhead that scales with the receiver mobility and the codebook resolution.

To improve existing search heuristics by decreasing the search space, we propose to exploit the spatial correlation between sub-6GHz and \ac{mmWave} bands, which has been observed by experimental measurements \cite{Nitsche_2015, Peter_2016, Hashemi_2017}.
Related techniques have been considered, for example, in \cite{Nitsche_2015}, where the authors use sub-6GHz spatial information to discover the \ac{mmWave} \ac{LoS} direction.
With the same goal, the authors in \cite{Ali_2017} have introduced a strategy to extract the spatial information from sub-6GHz, and they use it in \ac{mmWave} compressed beam-selection, assuming an analog architecture at the \ac{mmWave} band. 
Closer to the objective of this work, the study of deep learning methods for \ac{mmWave} beamforming, based on sub-6GHz channels information, is considered in \cite{Alrabeiah_2020}.
Although the proposed methods in \cite{Nitsche_2015, Ali_2017, Alrabeiah_2020} are reducing the beam training overhead, none of the proposed methods consider a \ac{HBF} architecture in the \ac{mmWave} band.
Moreover, the above-cited works are mainly limited to predict current \ac{mmWave} beams given current sub-6GHz channel measurements, but beam prediction from the spatially and temporally correlated sub-6GHz channel measurements have not been studied.

In contrast to previous studies, we consider important practical components in wireless communication systems, including \ac{NLOS} propagation and antenna polarization.
Moreover, the proposed method tracks future beams given past sub-6GHz channel measurements, which is a new aspect to the best of our knowledge.
The idea of predicting future beams from past sub-6GHz channel measurements is motivated by the spatial consistency of two close locations \cite{Kurras_2019}.
For example, with \ac{UE} mobility, the measured channels in two closely spaced \ac{UE} locations are often observed to have similar parameters such as the \ac{AoA}.
Hence, we use a neural network to exploit the sequence of sub-6GHz channel measurements, which are correlated in both spatial and time domain, to predict the \ac{mmWave} beam in the next time step(s).
In addition, we improve the prediction performance by fusing side-information such as the \ac{UE} location.
In particular, the proposed network design provides us with an opportunity to reduce the beam search space significantly if standard beam alignment strategies are employed.
We evaluate the performance of the proposed method with numerical simulations, and we compare it with the conventional search heuristics.
\begin{figure*}[!hb]
	\vspace{-12pt}
	\hrule
	\vspace{-5pt}
	\begin{equation}\label{eq.mut_inf_1}\tag{3}
		\resizebox{.95\hsize}{!}{$
			\begin{array} {rcl}
				\cbrace{ \FRFstar, \cbrace{ \FDKstar }_{\bar{k}=1}^{\bar{K}} } \in & \argmax\limits_{ \FRF, \cbrace{ \FDK }_{\bar{k}=1}^{\bar{K}}  } & \displaystyle \sum_{\bar{k}=1}^{\bar{K}} \log_2 \abs{ \mathbf{I} + \rho \Hmmk \FRF \FDK  \FDKH \FRFH \HmmkH } \vspace{-5pt}      \\
				                                                                   & \text{such that}                                                & \FRF \in       \begin{Bmatrix} \left.\begin{matrix} \recbrace{ {\FRFplus}^{\top} {\FRFminus}^{\top} }^{\top} \end{matrix}\right| 
				\FRFplus = \text{blkdiag} \nbrace{ \widehat{\mathbf{f}}_1, ..., \widehat{\mathbf{f}}_{N_{\text{RF}}} }, 
				\FRFminus = \text{blkdiag} \nbrace{ \widetilde{\mathbf{f}}_1, ..., \widetilde{\mathbf{f}}_{N_{\text{RF}}} }  \\ 
				\hspace{2pt} \text{and} \hspace{2pt}   \nbrace{\forall r \in \cbrace{1,...,N_{\text{RF}}}} \hspace{2pt} \widehat{\vect{f}}_r, \widetilde{\vect{f}}_r \in \mathcal{C}
				\end{Bmatrix}      ,         \\
				                                                                   &                                                                 & \displaystyle \sum_{\bar{k}=1}^{\bar{K}} \norm{ \FRF \FDK }^2 = \bar{K}N_s.                                                      
			\end{array}
		$}
	\end{equation}
	\vspace{-5pt}
	\begin{equation}\label{eq.mut_inf_2}\tag{4}
		\resizebox{0.95\hsize}{!}{$
			\begin{array} {rcl}
				\hspace{74pt}  \FRFstar \in & \argmax\limits_{ \FRF  } & \displaystyle \sum_{\bar{k}=1}^{\bar{K}} \log_2 \abs{ \mathbf{I} + \rho \Hmmk \FRF \nbrace{\FRF^H \FRF}^{-1} \FRFH \HmmkH } \vspace{-5pt} \\
				                            & \text{such that}         & \FRF \in       \begin{Bmatrix} \left.\begin{matrix} \recbrace{ {\FRFplus}^{\top} {\FRFminus}^{\top} }^{\top} \end{matrix}\right|          
				\FRFplus = \text{blkdiag} \nbrace{ \widehat{\mathbf{f}}_1, ..., \widehat{\mathbf{f}}_{N_{\text{RF}}} } ,
				\FRFminus = \text{blkdiag} \nbrace{ \widetilde{\mathbf{f}}_1, ..., \widetilde{\mathbf{f}}_{N_{\text{RF}}} }  \\ 
				\hspace{2pt} \text{and} \hspace{2pt}   \nbrace{\forall r \in \cbrace{1,...,N_{\text{RF}}}} \hspace{2pt} \widehat{\vect{f}}_r, \widetilde{\vect{f}}_r \in \mathcal{C}
				\end{Bmatrix}.
			\end{array}
		$}
	\end{equation}
\end{figure*}
\vspace{-8pt}
\section{System Model and Problem Statement}
\label{sec:format}
\vspace{-7pt}
\subsection{System Model}
\label{system_model}
\vspace{-5pt}
In the following, we consider a dual-band wireless \ac{MIMO}-\ac{OFDM} communication system operating in both sub-6GHz and \ac{mmWave} bands, and consisting of single \ac{BS} and single \ac{UE}.
The \ac{BS} is assumed to employ a \ac{UPA} with $N_{\text{tx}}^{\text{sub-6}}$ and $N_{\text{tx}}$ antenna elements with cross-polarization in the sub-6GHz and \ac{mmWave} bands, respectively.
The \ac{UE} is assumed to employ a single antenna in sub-6GHz and a \ac{UPA} with $N_{\text{rx}}$ antenna elements in \ac{mmWave} band.
The total number of \ac{OFDM} subcarriers in the sub-6GHz and \ac{mmWave} bands are denoted by $K$ and $\bar{K}$, respectively.
In this work, we exploit the uplink channel measurements in the sub-6GHz band to assist the beamforming in the downlink \ac{mmWave} band. 
With cross-polarization, the uplink sub-6GHz channel at subcarrier $k \in \cbrace{1,...,K}$ and the downlink \ac{mmWave} channel at subcarrier $\bar{k} \in \cbrace{1,...,\bar{K}}$ are denoted by $\vect{h}\recbrace{k} \in \mathbb{C}^{2N_{\text{tx}}^{\text{sub-6}}}$ and $\Hmmk \in \mathbb{C}^{2N_{\text{rx}} \times 2N_{\text{tx}}}$, respectively.
We assume that the \ac{UE} can perform optimal decoding from a received signal in the \ac{mmWave} band with fully digital hardware, and we focus on the hybrid precoding design at the \ac{BS}.
An illustration of the hybrid precoding architecture in the \ac{mmWave} band with $N_s$ data streams, $N_{\text{RF}}$ RF chains, and $N_{\text{tx}}$ transmit antennas with cross-polarization is given in Fig.~\ref{fig.hybrid_precoding_architecture}.
Let $\FRF \in \complex^{2 N_{\text{tx}} \times N_{\text{RF}}}$ be an RF precoding matrix, and $\FDK \in \complex^{ N_{\text{RF}} \times N_s}$ be a digital baseband precoding matrix at the $\bar{k}^{\text{th}}$ OFDM subcarrier in the \ac{mmWave} band.  
The RF precoder is decomposed into a $+45^{\circ}$ polarized precoder $\FRFplus \in \complex^{N_{\text{tx}} \times N_{\text{RF}}}$ and a $-45^{\circ}$ polarized precoder $\FRFminus \in \complex^{N_{\text{tx}} \times N_{\text{RF}}} $, and we define $\FRF := \recbrace{ \FRFplus^{\top} \FRFminus^{\top}}^{\top}$.
The received signal at subcarrier $\bar{k}$ for a transmitted symbol $\vect{s}\recbrace{\bar{k}} \in \complex^{N_s}$ is given by
\begin{equation}
	\vect{y}\recbrace{\bar{k}} = \Hmmk \FRF \FDK \vect{s}\recbrace{\bar{k}} + \vect{n}\recbrace{\bar{k}},
\end{equation}
where $\vect{n}\recbrace{\bar{k}} \in \complex^{2 N_{\text{rx}}}$ denotes the additive white Gaussian noise.
The cross-polarized \ac{mmWave} channel at subcarrier $\bar{k}$ is represented in block form as
\begin{equation}
	\Hmmk = \begin{bmatrix}
	\Hmmk_{+45^{\circ}} & \Hmmk_{\pm 45^{\circ}}\\ 
	\Hmmk_{\mp 45^{\circ}} & \Hmmk_{-45^{\circ}}
	\end{bmatrix},
\end{equation}
\noindent where the diagonal blocks $\Hmmk_{+45^{\circ}}$ and $\Hmmk_{-45^{\circ}}$ represent the co-polarized, and the off-diagonal blocks $\Hmmk_{\pm 45^{\circ}}$ and $\Hmmk_{\mp 45^{\circ}}$ represent the cross-polarized components.
In this work, we consider a hybrid precoding design with fixed subarray architecture, which means that each RF chain is connected to one of the non-overlapping subsets of antenna elements (see Fig.~\ref{fig.hybrid_precoding_architecture}).
For simplicity, we assume that all RF chains have the same subset size.
With this architecture, the analog precoding matrix with $+45^{\circ}$ polarization takes the form of a block diagonal matrix as follows:
\begin{figure}[!t]
	\centering
	\centerline{\includegraphics[width=77.0mm]{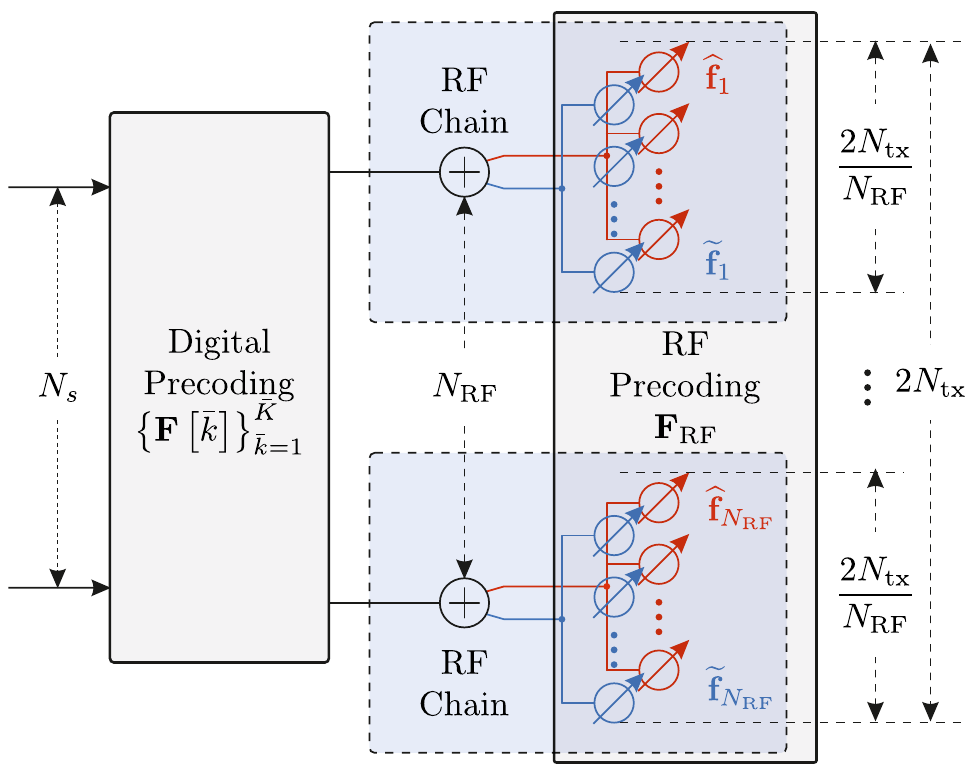} \vspace{-10pt}}
	\caption{Hybrid analog/digital precoding architecture in a \ac{mmWave} band \ac{MIMO}-\ac{OFDM} system with cross-polarization.}
	\label{fig.hybrid_precoding_architecture}
	\vspace{-20pt}
\end{figure}
\begin{equation*}
	\FRFplus = \blkdiag{\widehat{\vect{f}}_1,..., \widehat{\vect{f}}_{N_{\text{RF}}}} = \begin{bmatrix}
	\widehat{\vect{f}}_1 & \hdots & \vect{0}\\ 
	\vdots & \ddots  & \vdots\\ 
	\vect{0} & \hdots & \widehat{\vect{f}}_{N_{\text{RF}}}
	\end{bmatrix},
\end{equation*}
where $\nbrace{\forall r \in \cbrace{1,...,N_{\text{RF}} } } \hspace{2pt} \widehat{\vect{f}}_{r}$ are design parameters.
\begin{figure*}[!t]
	\centering
	\centerline{\includegraphics{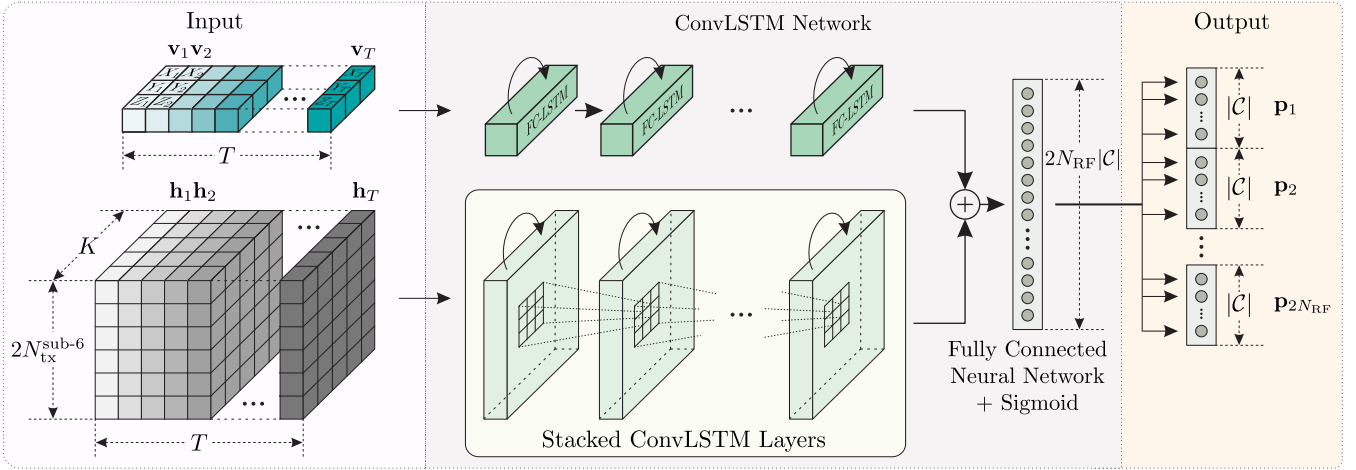}}
	\vspace{-8pt}
	\caption{The proposed neural network architecture for beam prediction/tracking. The network takes the spatiotemporal sub-6GHz channel measurements with $T$ samples in sequence and predicts the \ac{mmWave} beam indices in $T+1$ time step.}
	\label{neural_network_diagram}
	\vspace{-10pt}
\end{figure*}
\noindent The RF precoding matrix with $-45^{\circ}$ polarization is defined in a similar way; i.e., $\FRFminus = \blkdiag{\widetilde{\vect{f}}_1,..., \widetilde{\vect{f}}_{N_{\text{RF}}}}$.
In the above, $\nbrace{\forall r \in \cbrace{1,...,N_{\text{RF}}} }$ $\widehat{\vect{f}}_r, \widetilde{\vect{f}}_r \in \complex^{N_{\text{tx}}/N_{\text{RF}}}$ is a beamforming vector associated with the $r^{\text{th}}$ RF chain, and we assume that each beamforming vector is selected from a predefined finite codebook $\mathcal{C}$, i.e., $\nbrace{\forall r \in \cbrace{1,...,N_{\text{RF}}} }$ $\widehat{\vect{f}}_r, \widetilde{\vect{f}}_r \in \mathcal{C}$ \cite{Lee2011}.
\vspace{-10pt}
\subsection{Problem Statement}
\vspace{-5pt}
Assuming that the symbols are sampled from a Gaussian distribution, i.e., $\vect{s}\recbrace{\bar{k}} \sim \mathcal{N} \nbrace{\vect{0},\vect{I}}$ \cite{Goldsmith_2003, El_2014}, a general approach for hybrid precoding is to maximize the mutual information given in \eqref{eq.mut_inf_1} at the bottom of this page, where $\rho$ denotes the \ac{SNR}.
Optimizing \eqref{eq.mut_inf_1} directly is challenging due to (i) the non-convex constraint on $\FRF$, and (ii) the coupling between the analog and digital matrices, which arises in the power constraint.
Alternatively, it is shown in \cite{Alkhateeb_2016, Park_2017} that the optimal digital precoder can be written as a function of the optimal RF precoders $\FRFstar$, e.g., $\nbrace{ \forall \bar{k} \in \cbrace{1,...,\bar{K}} } \FDKstar = f \nbrace{\FRFstar}$ where $f: \complex^{2 N_{\text{tx}} \times N_{\text{RF}}} \rightarrow \complex^{ N_{\text{RF}} \times N_s}$ is a known function given in \cite{Alkhateeb_2016, Park_2017}. 
This enables us to focus on the optimization of the RF precoder $\FRFstar$, since the digital precoder $\nbrace{\forall \bar{k} \in \cbrace{1,...,\bar{K}}} \hspace{2pt} \FDKstar$ can be easily recovered from $\FRFstar$ with the function $f$.
Using the relation between digital and analog precoders, i.e., $\nbrace{ \forall \bar{k} \in \cbrace{1,...,\bar{K}} } \FDKstar = f \nbrace{\FRFstar}$, we can remove the digital precoders from \eqref{eq.mut_inf_1}.
Effectively, this leads to the new problem in \eqref{eq.mut_inf_2} where the mutual information-based hybrid precoding is determined only by the RF precoders.

Although the formulation in \eqref{eq.mut_inf_2} simplifies the hybrid precoding problem, the optimization is still hard owing to the discrete constraints imposed on the analog precoder $\FRF$.
In principle, the problem in \eqref{eq.mut_inf_2} could be solved via exhaustive search.
Performing this search, however, requires either estimating the \ac{mmWave} channel $\vect{H}$ or an online exhaustive beam training, but both approaches have a large signalling overhead.
In the following, we propose a deep learning-based method that predicts the RF precoders from sub-6GHz channel measurements.
\vspace{-10pt}
\section{Deep Learning-Based Hybrid Precoding}
\vspace{-7pt}
\setcounter{equation}{4}
To cope with the problem in \eqref{eq.mut_inf_2}, we assume that there exists an ideal mapping $\Xi$ that maps the sequence of $T$ sub-6GHz channel measurements $\recbrace{\mathbf{h}_1, ..., \mathbf{h}_T}$ in the uplink to some output $\recbrace{\mathbf{p}^{\star}_1,...,\mathbf{p}^{\star}_{2N_{\text{RF}}}}^{\top}$ (to be described later) such that the downlink \ac{mmWave} RF precoder $\FRFstar$ can be reconstructed from $\recbrace{\mathbf{p}^{\star}_1,...,\mathbf{p}^{\star}_{2N_{\text{RF}}}}^{\top}$, and the reconstructed $\FRFstar$ maximizes the objective given in \eqref{eq.mut_inf_2}.
Formally, the ideal mapping $\Xi$ is defined as follows
\begin{equation}\label{eq.mapping_Xi}
	\resizebox{0.91\hsize}{!}{$
			\Xi: \complex^{2N_{\text{tx}}^{\text{sub-6}} \times K \times T}  \rightarrow \mathbb{R}^{2 N_{\text{RF}} \abs{\mathcal{C}}}: 	\recbrace{\mathbf{h}_1, ..., \mathbf{h}_T} \mapsto \recbrace{\mathbf{p}^{\star}_1,...,\mathbf{p}^{\star}_{2N_{\text{RF}}}}^{\top}.
			$}
\end{equation}

In the following, we explain the output design of the ideal mapping $\Xi$ and describe the reconstruction mechanism of $\FRFstar$ from $\recbrace{\mathbf{p}^{\star}_1,...,\mathbf{p}^{\star}_{2N_{\text{RF}}}}^{\top}$.
First, we assume that the output indices $r \in \cbrace{1,...,N_{\text{RF}}}$ and $r \in \cbrace{N_{\text{RF}}+1,...,2N_{\text{RF}}}$ refer to the $+45^{\circ}$ and $-45^{\circ}$ polarization, respectively.
Second, we assume that the elements of the vector $\vect{p}^{\star}_r$  are one-hot encoded, i.e., $\left (  \forall r \in \cbrace{1,..., 2 N_{\text{RF}}} \right )$ $\vect{p}^{\star}_r \in \left \{  0,1\right \} ^{\abs{\mathcal{C}}}$, where $|\mathcal{C}|$ indicates the size of a predefined codebook $\mathcal{C}$. 
In this way, the indices of the vector $\nbrace{\forall r \in \cbrace{1,..., 2 N_{\text{RF}}}} \hspace{2pt} \vect{p}^{\star}_r$ are associated with the respective codeword indices from a finite codebook $\mathcal{C}$.
Now, to relate the selected optimal beam indices (in the sense of maximizing the objective in \eqref{eq.mut_inf_2}) from a codebook $\mathcal{C}$ for the respective RF chain with the corresponding polarization to the output of the ideal mapping $\Xi$, we further assume that the each vector  $\nbrace{\forall r \in \cbrace{1,..., 2 N_{\text{RF}}}} \hspace{2pt} \vect{p}^{\star}_r$ contains the value one at the index corresponding to the selected optimal beam index, and zeros elsewhere. 

With the output design explained above, the reconstruction of the precoder $\FRFstar$ from $\recbrace{\vect{p}^{\star}_1,..., \vect{p}^{\star}_{2N_{\text{RF}}}}^{\top}$ is performed as follows.  
Once we the obtain the output $\recbrace{\vect{p}^{\star}_1,..., \vect{p}^{\star}_{2N_{\text{RF}}}}^{\top}$ from $\Xi$, we construct the precoding matrices $\FRFplus^{\star}$ and $\FRFminus^{\star}$ via
\begin{equation}\label{eq.optimal_precod_Xi}
	\resizebox{1.00\hsize}{!}{$
		\FRFplus^{\star} = \blkdiag{ \widehat{\vect{f}}_1^{\star},...,\widehat{\vect{f}}_{N_{\text{RF}}}^{\star} }, \nbrace{\forall r \in \cbrace{1,...,N_{\text{RF}}}} \begin{array}{l}
		\widehat{\vect{f}}_r^{\star} = \vect{c}_{i^{\star}_r} \in \mathcal{C}\\ 
		i^{\star}_r \in \underset{i = 1,...,\abs{\mathcal{C}}}{\text{arg max}} \nbrace{\recbrace{\vect{p}^{\star}_r}_i}
		\end{array},
	$} 
\end{equation}
\begin{equation}\label{eq.optimal_precod2_Xi}
	\resizebox{1.00\hsize}{!}{$
		\FRFminus^{\star} = \blkdiag{ \widetilde{\vect{f}}_1^{\star},...,\widetilde{\vect{f}}_{N_{\text{RF}}}^{\star} }, \nbrace{\forall r \in \cbrace{1,...,N_{\text{RF}}}} \begin{array}{l}
		\widetilde{\vect{f}}_r^{\star} = \vect{c}_{i^{\star}_r} \in \mathcal{C}\\ 
		i^{\star}_r \in \underset{i = 1,...,\abs{\mathcal{C}}}{\text{arg max}} \nbrace{\recbrace{\vect{p}^{\star}_{r+N_{\text{RF}}}}_i}
		\end{array},
	$}
\end{equation}
where $\vect{c}_{i^{\star}_r}$ represents the selected optimal codeword from a finite codebook $\mathcal{C}$ by the $r^{\text{th}}$ RF chain with corresponding polarization.
Further, the precoding matrix $\FRFstar$ that maximizes the objective in \eqref{eq.mut_inf_2} is constructed as $\FRFstar = \recbrace{ \FRFplus^{\star^{\top}} \FRFminus^{\star^{\top}}}^{\top}$.
Although the ideal mapping $\Xi$ solves \eqref{eq.mut_inf_2} by assumption, it is challenging to analytically characterize it.
Therefore we propose a deep neural network that learns an ideal mapping $\Xi$ from data. 

The proposed neural network architecture is given in Fig.~\ref{neural_network_diagram}.
With the setting that the RF precoders are taken from a finite codebook, we consider the beam indices as labels and pose the RF precoding problem in \eqref{eq.mut_inf_2} as multi-label classification \cite{Tsoumakas_2007}.
The labels are constructed as follows.
Given a finite codebook $\mathcal{C}$ and a tuple $\nbrace{ \recbrace{\vect{h}_1, ..., \vect{h}_T} , \vect{H}_{T+1}}$, where $\recbrace{\vect{h}_1, ..., \vect{h}_T}$ and $\vect{H}_{T+1}$ are the $T$ sub-6GHz channel measurements in sequence and the \ac{mmWave} channel at time step $T+1$, respectively, we perform an exhaustive search following \eqref{eq.mut_inf_2} with $\vect{H}_{T+1}$ to obtain $\FRFstar$.
Next, we decompose $\FRFstar = \recbrace{ \FRFplus^{\star^{\top}} \FRFminus^{\star^{\top}}}^{\top}$ and retrieve $\nbrace{\forall r \in \cbrace{1,..., N_{\text{RF}}}} \hspace{2pt} i^{\star}_r$ by following the rules in \eqref{eq.optimal_precod_Xi} and \eqref{eq.optimal_precod2_Xi}.
Further, with $i^{\star}_r$ we construct the one-hot encoded ground truth labels $\recbrace{\vect{p}^{\star}_1,..., \vect{p}^{\star}_{2N_{\text{RF}}}}^{\top}$.

Given a sequence $\recbrace{\vect{h}_1, ..., \vect{h}_T}$, we extract both the spatially and temporally correlated features from sub-6GHz channel measurements by utilizing \ac{ConvLSTM} layers. 
The \ac{ConvLSTM} layers have a convolutional structure in both the input-to-state and state-to-state transitions.
To project the extracted features onto $2 N_{\text{RF}} \abs{\mathcal{C}}$-dimensional vector $\recbrace{\vect{p}_1,..., \vect{p}_{2N_{\text{RF}}}}^{\top}$, we customize the task-specific layer based on a fully connected neural network with sigmoid activations so that $\left (  \forall r \in \cbrace{1,..., 2 N_{\text{RF}}} \right )$ $\vect{p}_r \in \recbrace{  0,1} ^{\abs{\mathcal{C}}}$.
With this design, the output of neural network indicates beam priorities.
In other words, the index of the first highest value in $\left (  \forall r \in \cbrace{1,..., 2 N_{\text{RF}}} \right )$ $\vect{p}_r$ is the best suggested beam from the predefined codebook $\mathcal{C}$. 
Accordingly, the indices of the second, third and $n^{\text{th}}$ highest values, with $n \le \abs{\mathcal{C}}$, in $\left (  \forall r \in \cbrace{1,..., 2 N_{\text{RF}}} \right )$ $\vect{p}_r$ are the second, third, and $n^{\text{th}}$ best suggestions, respectively.
We train a neural network by trying to minimize with the stochastic gradient method the expected value of the binary cross-entropy loss \cite{Tsoumakas_2007} given by
\begin{equation}
	\resizebox{1.0\hsize}{!}{$
		\begin{array}{c}
			l \nbrace{ \recbrace{\vect{p}_1^{\star},...,\vect{p}_{2N_{\text{RF}}}^{\star}}^{\top}, \recbrace{\vect{p}_1,...,\vect{p}_{2N_{\text{RF}}}}^{\top} } =                                                                                                                                                       \\
			= -\dfrac{1}{2 N_{\text{RF}} \abs{\mathcal{C}}} \displaystyle \sum_{r = 1}^{2 N_{\text{RF}}} \sum_{i = 1}^{\abs{\mathcal{C}}} \Big( \recbrace{\vect{p}_r^{\star}}_i \log \nbrace{\recbrace{\vect{p}_r}_i} + \nbrace{1 - \recbrace{\vect{p}_r^{\star}}_i }  \log \nbrace{\recbrace{1 - \vect{p}_r}_i} \Big). 
		\end{array}
	$}
\end{equation}
In addition to sub-6GHz channel measurements, we also fuse side information to improve the network performance. 
In more detail, we assume that the \ac{UE} location information (e.g., Cartesian coordinates) can be independently acquired by the \ac{BS} with the use of available positioning technologies.
Similar to the sequence of sub-6GHz channel measurements, we construct the sequence with \ac{UE} location information, i.e., $\recbrace{\vect{v}_1,...,\vect{v}_T}$, and use a tuple $\nbrace{ \recbrace{\vect{h}_1, ..., \vect{h}_T}, \recbrace{\vect{v}_1,...,\vect{v}_T} }$ as an input to a neural as show in Fig.~\ref{neural_network_diagram}.  
To take the advantage of \ac{UE} location information, we use the \ac{FC-LSTM} layers.
\vspace{-10pt}
\section{Numerical Results}
\vspace{-5pt}
\subsection{Performance Evaluation Metrics}
\vspace{-5pt}
Owing to the fact that the proposed network produces vectors with the elements ranging between 0 and 1, we evaluate the network by measuring the frequency at which the neural network correctly predicts the labels within its \textit{best-n} predictions.
The \textit{best-n} prediction accuracy is denoted by $A_{\text{best-n}}$ and it is defined by
\vspace{-3pt}
\begin{equation}
	A_{\text{best-n}} = \dfrac{1}{S} \displaystyle \sum_{s=1}^S \mathbbm{1}^n_{ \nbrace{ \recbrace{ \vect{p}_1^n,..., \vect{p}^n_{2N_{\text{RF}}} },  \recbrace{ \vect{p}_1^{\star},..., \vect{p}^{\star}_{2N_{\text{RF}}} }}  },
\end{equation}
where $S$ is the test dataset size, and $\mathbbm{1}_{\nbrace{\cdot,\cdot}}$ is the indicator function given by
\vspace{-5pt}
\begin{equation}
	\vspace{-2pt}
	\resizebox{1.0\hsize}{!}{$
		\mathbbm{1}^n_{ \nbrace{ \recbrace{ \vect{p}_1^n,..., \vect{p}^n_{2N_{\text{RF}}} },  \recbrace{ \vect{p}_1^{\star},..., \vect{p}^{\star}_{2N_{\text{RF}}} }}  }:= \left\{\begin{array}{ll}
		1 &, \text{if}  \displaystyle \sum_{r=1}^{2N_{\text{RF}}} \sum_{i=1}^{\abs{\mathcal{C}}} \recbrace{ \vect{p}_r^n }_i \recbrace{\vect{p}_r^{\star}}_i = 2N_{\text{RF}}, \\ 
		0 &, \text{otherwise.}
		\end{array}\right.
	$}
\end{equation}
The vector $\vect{p}_r^n$ represents the predicted labels with the \textit{best-n} criterion, and it contains ones in the indices with the $n$ largest values of $\vect{p}_r$, and zeros elsewhere.
In addition to the \textit{best-n} prediction accuracy, we also evaluate the performance of the proposed method in terms of the spectral efficiency with the predicted \ac{mmWave} beams.
\vspace{-10pt}
\subsection{Simulation}
\vspace{-5pt}
To evaluate the performance of the proposed methods, we generate spatially and time correlated channels with \ac{QuaDRiGa} \cite{Jaeckel_2016}.
The input sequence to the network consists of $T=5$ spatiotemporal channel measurements, and the proposed method predicts the \ac{mmWave} beams in $T+1$ time steps.
The finite codebook is adopted from \cite{Xie_2013}.
For network training we use three ConvLSTM layers with the kernel size of $10 \times 10$, and the training and testing datasets consist of 95K and 19K samples, respectively.
The learning rate, number of epochs and batch size are 0.001, 10 and 500, respectively.
The remaining simulation parameters are given in Table~\ref{tab:1}.
{\renewcommand{\arraystretch}{1.3}
	\begin{table}[!b]
		\vspace{-15pt}
		\resizebox{1.00\hsize}{!}{$
			\begin{tabular}{|c|c|c|}
				\hline
				\multirow{2}{*}{\textbf{Parameter}} & \multicolumn{2}{c|}{\textbf{Transceiver}}                                                             \\ \cline{2-3} 
				                            & \textbf{sub-6GHz} & \textbf{mmWave}                              \\ \hline
				Carrier frequency {[}GHz{]} & 3.6               & 26                                           \\ \hline
				Bandwidth {[}MHz{]}         & 20                & 800                                          \\ \hline
				OFDM subcarriers            & $K = 32$          & $\bar{K} = 512$                              \\ \hline
				BS antenna size {[}UPA{]}   & $4 \times 4$      & $8 \times 8$                                 \\ \hline
				UE anetnna size             & 1                 & $2 \times 2$ UPA                             \\ \hline
				Polarization                        & \multicolumn{2}{c|}{$\pm 45^{\circ}$}                                                                     \\ \hline
				Signal processing           & Fully Digital     & \begin{tabular}[c]{@{}c@{}}HBF with subarray \\ structure $(N_{\text{RF}}=2)$\end{tabular} \\ \hline
				Propagation scenario                & \multicolumn{2}{c|}{3GPP\_38.901\_UMa\_NLOS \cite{3gpp_2017_}}                                                                             \\ \hline
				UE mobility                         & \multicolumn{2}{c|}{30 km/h}                                                                \\ \hline
			\end{tabular}
		$}
		\caption{Simulation parameters}
		\label{tab:1}
	\end{table}}
Although the main goal of the proposed method is to skip channel estimation in the \ac{mmWave} band, as a comparison, we also present the results of the proposed method trained with \ac{mmWave} channels. The performance of the proposed network trained with sub-6GHz and \ac{mmWave} channels is denoted by \texttt{ConvLSTM} and $\overline{\texttt{ConvLSTM}}$, respectively.
The superscript $^{\text{loc}}$ and the subscripts $_{1,3,5}$ denote the performance with location information and with \textit{best-1,3,5} beam predictions, respectively.

\begin{figure}[!ht]
	\centering
	\centerline{\includegraphics[width=84.5mm]{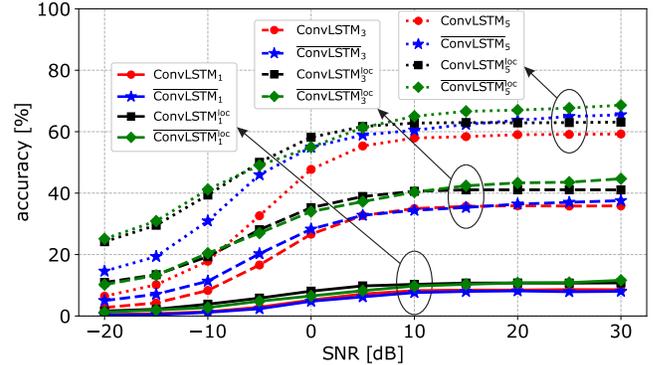}}
	(a) Prediction accuracy
	\centerline{\includegraphics[width=84.5mm]{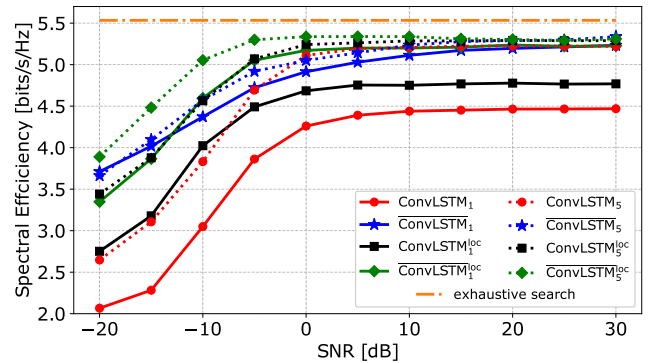}}
	(b) Spectral efficiency
	\vspace{-7pt}
	\caption{Performance of the proposed method.}
	\label{fig.sim_result}
	\vspace{-10pt}
\end{figure}

In Fig.~\ref{fig.sim_result}(a), we compare the prediction accuracy of the proposed method with \textit{best-n} criterion (e.g., $ n = 1,3,5$) with and without side information.
The results in Fig.~\ref{fig.sim_result}(a) experimentally justify the advantage of fusing side information.
In particular, utilization of side information with highly noisy channel measurements improves the performance noticeably.

In Fig.\ref{fig.sim_result}(b), we provide the spectral efficiency with the beams predicted by the proposed model. 
The x-axis values represent the \ac{SNR} of input channel measurements to proposed network and the \ac{SNR} of \ac{mmWave} channels, which we utilize to to compute the spectral efficiency, is fixed to 30dB.
Given $N_{\text{RF}}=2$ and $\abs{\mathcal{C}} = 32$, the exhaustive search method in Fig.\ref{fig.sim_result}(b) evaluates $\abs{\mathcal{C}}^{2N_{\text{RF}}} =$ 1,048,576 configurations.
In comparison, for $n=5$, the prediction with \textit{best-n} criterion reduces the search space to $625$ configurations. 

The results in Fig.~\ref{fig.sim_result} demonstrate that the proposed method provides reasonable achievable rates while significantly reducing the signalling overhead. 
Moreover, exploiting side information can increase the system performance in terms of both prediction accuracy and spectral efficiency.
\vspace{-10pt}
\vspace{-10pt}
\section{Conclusion}
\vspace{-7pt}
In this work, we proposed a deep learning-based hybrid precoding scheme in the \ac{mmWave} band.
The proposed method predicts the \ac{mmWave} beams by exploiting the spatiotemporal correlation between the sub-6GHz and \ac{mmWave} bands.
Simulations showed that the proposed method can significantly reduce the signalling overhead in \ac{mmWave} band with hybrid architectures by maintaining good spectral efficiency in the system.
Moreover, we showed that side information such as the location of the \ac{UE} can significantly improve the system performance. 

\clearpage
\bibliographystyle{IEEEbib}
\bibliography{strings,refs}
\end{document}